# Local field enhancement using slanted Bound States in the Continuum cavities and plasmonic nanoantenna

Liyi Hsu,[1] Fadi I. Baida[2] and Abdoulaye Ndao[1,*]

[1]Department of Electrical and Computer Engineering & Photonics Center, Boston University, 8 Saint Mary's Street, Boston, Massachusetts 02215, USA

[2]Institut FEMTO-ST, UMR 6174 CNRS, Departement d'Optique P. M. Duffieux, Universite de Franche–Comte, 25030 Besancon Cedex, France

*Corresponding author: andao@bu.edu

**Abstract**

Over the last few years, optical nanoantennas are continuously attracting interest owing to their ability to efficiently confine, localize resonance, and significantly enhanced electromagnetic fields at subwavelength scale. However, such strong confinement can be further enhanced by using an appropriate combination of optical nanoantenna and Slanted Bound states in the continuum cavities. Here, we propose to synergistically bridge the plasmonic nanoantenna and high optical quality-factor cavities to numerically demonstrate three orders of magnitude local intensity enhancement. The proposed hybrid system paves a new way for applications requiring highly confined fields such as optical sensing, nonlinear optics, quantum optics, etc.

**Introduction**

High optical quality-factors cavities induced strong local field enhancement are in tremendous demand in nano-optics due to the large number of applications requiring strong light-matter interaction such as biosensors, compact spectral splitting solar cells, low-threshold lasers, and single-photon sources. In the last decade, many different approaches have been proposed to combine high quality-factors and subwavelength confinement using hybrid photonic-plasmonic structure [1-9]. However, usually, large-quality factors come at the expense of compactness. All those traditional approaches have been stymied in their effort to break this compromise because they all rely on the same core principle that, to increase the radiation quality factor, no outgoing waves must be allowed. To break free of this paradigm, it is necessary to bring a new perspective. Bound states in the Continuum (BICs) are originally proposed in a very different field of wave physics, namely quantum mechanics [10-13] to achieve high-quality factors. BICs are states that are bound, i.e., have no radiation losses, even though coupling to the radiation continuum is, strictly speaking, allowed. They offer the promise of simultaneously large quality factors, compact devices [14-22], and local field enhancement, defined as the ratio of the total intensity divided by the intensity of the incident field.

On the other hand, optical nanoantenna (NA), have been of great interest because of their ability to efficiently confine, localize resonance, and significantly enhance electromagnetic fields at subwavelength scale [23-26].

Such strong confinement can be further enhanced by using an appropriate combination between optical NA and Bound states in the continuum. Here, we combined the high Q-factor of the BIC and the ability of the plasmonic NA to confine light at subwavelength scale and demonstrate an enhancement of three orders of magnitude larger than what can be obtained within a single resonator. This versatile platform ensures the ultra-small mode volume from the NA and the high-quality factor from the BIC cavities, thus avoiding a strong coupling that can break the resonances and leading to large field enhancement. It paves a new way for applications requiring highly confined fields such as optical sensing, nonlinear optics, quantum optics, etc.

In this paper, we propose to use a one-dimension slanted Bound States in the continuum cavities (SBIC) at an angle $\alpha$ and plasmonic optical nanoantennas as a building block (see Fig. 1). The SBIC consists of two high-index ridges (Si), separated by a narrow low-index gap (water,) deposited on a $SiO_2$ substrate. First, we start with a symmetrical vertical grating (a=0°) and go towards a slanted grating ($\alpha$) with symmetry breaking [27-29]. Second, we discuss the plasmonic NA design. After that, the hybrid system is designed and simulated to demonstrate a strong field intensity enhancement.

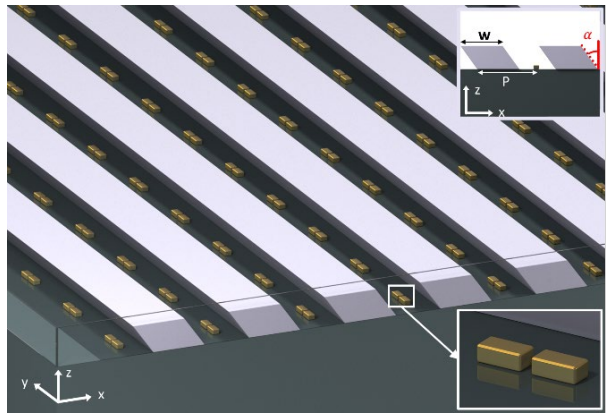

Fig. 1 Schematic of a hybrid SBIC-NA, which is made of slanted Bound States in the continuum Cavities (SBIC) and plasmonic nanoantenna (NA). The grating consists of two high-index ridges (Si), separated by a narrow low-index gap (water). The NA is composed of gold nanoparticles with rectangular shapes. The incident wave is TE polarized with the electric field E along the y-direction impinging the grating at normal incidence.

**Design of the Bound state in the Continuum cavities**

To analyze our proposed system, we first simulated a high refractive index vertical ($\alpha$=0°) grating (Si, n = 3.47) sitting on low-index gap ($SiO_2$, n=1.46) and surrounded by water (n=1.33). The period (p) of the grating array is 500nm, the Si slab width (w) is 225 nm and the height is 600nm. The structure is illuminated along angle θ varying from 0° to 55° with polarization along the axis of the dielectric grooves (y-polarization). To design our system, we performed numerical simulations using a 3D-FDTD homemade code integrating a Critical Points model to accurately consider the dispersion of the material [30].

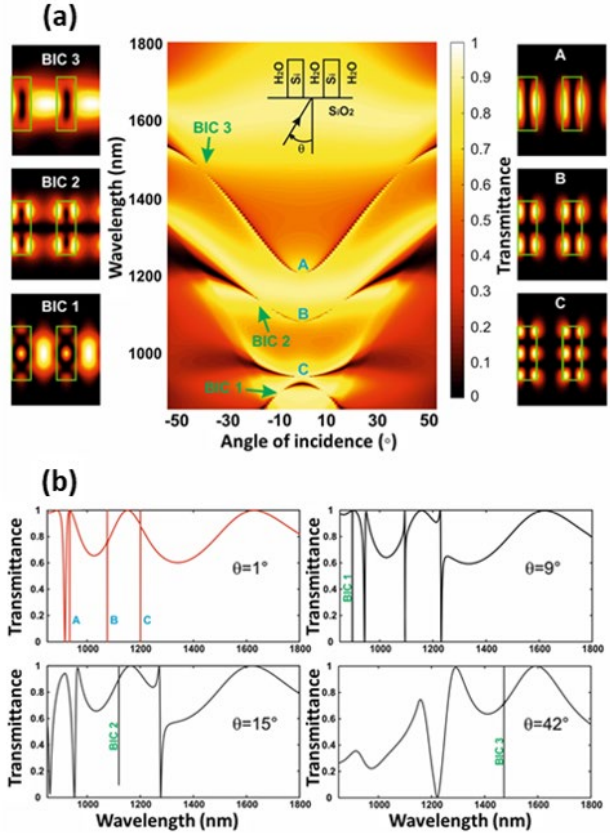

Fig. 2 (a) Transmission spectra for the different incident angles θ. The inset diagram is a schematic of the modeled structure. The subfigures on the right give the electric intensity distributions at the vicinity of the grating near the BICs (A, B, and C) at θ=1° at λ=934.43nm, λ=1076.21nm, and λ=1200.88 nm respectively. The subfigures on the left correspond to the same distribution near the BICs that occur at oblique incidence (BIC 1 at θ=9° and λ=895.05 nm, BIC 2 at θ =15° and λ =1124.38 nm and BIC 3 at θ =42° and λ =1474.09 nm).

Diagram of Fig. 2 shows the transmission coefficient, in which the quality factor of the several modes (A, B, C at normal incidence and BIC 1, BIC 2, and BIC 3 at oblique incidence) can be seen to tend slowly to infinity from their vanishing linewidths. Those vanishing linewidths indicate a trapped state with no leakage named Bound States in the continuum. To Substantiate the existence of BICs, we calculated the electric field intensity distributions for all the BICs and determine those for which the enhancement occurs in water. In fact, the NA should be placed in contact with water to be sensitive to its optical index variations induced by the presence of biological molecules. As shown in Fig. 2, only BIC 1 and BIC 3, excited at oblique incidence of θ=9.2° and θ=41.5° at λ=893.05nm and λ=1471.09nm respectively, exhibit light confinement in the groove central zone (water). Nevertheless, for the sake of compactness and integration (at the end of an optical fiber for instance), it is more convenient to work at normal incidence. To this end, we propose to bring these two BICs at θ=0° by tilting the Si slabs, and thus, by breaking the mirror symmetry of the grating [27-29].

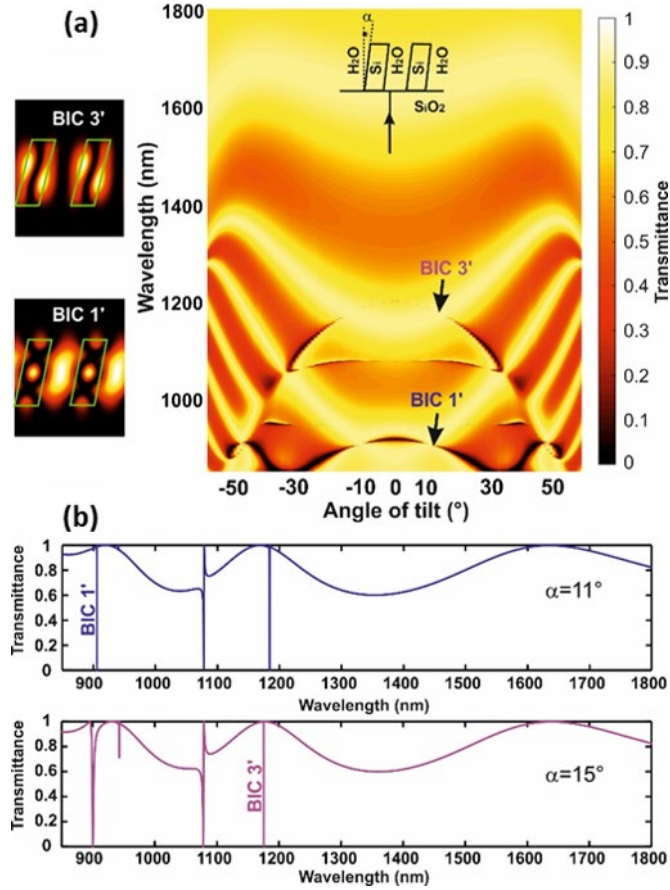

Fig. 3 (a) Transmission spectra with different slanted angles of the grating structure. The two sub-figures at the left depict the electric intensity distribution around the grating for the two resonances denoted by BIC 1' and BIC 3'. (b) Transmittance spectra at α=11° (left) and α=15° (right) near the BIC 1' and BIC 3' respectively.

This symmetry breaking results in two fundamental advantages. The first one consists of simplifying experiment conditions and the second one to efficiently couple the fundamental mode of the NA to the grating in view of high field enhancement. Transmission spectra at normal incidence for tilt angle α varying from -60° to 60° are depicted on figure 3 where one can see that the BIC 1 (new BIC 1') is now slightly shifted toward the red spectral region (λ=905.83nm) for α = 12° while the BIC 3 (new BIC 3') appears at λ=1175.3 nm for α=15°.

Unfortunately, the intensity distribution of these two modes, presented at the left of Fig. 3(a), shows that BIC 3' does not match an electric field confinement in the groove region. The only BIC 1' ($Q \approx 10^{12}$) remains compatible with a normal incident (see the inset at the bottom left of Fig. 3(a)) in addition having the smaller tilt angle (α=12°).

**Plasmonic NA design**

Next, we designed a nanoantenna that has a resonance frequency close to that of the BIC 3'. The NA is made in gold and the considered permittivity of the gold is taken from [31] and adapted through a Drude Critical Points (DCP) Model integrated into the FDTD codes. Using commercial software CST and homemade FDTD codes, we performed the numerical simulation, considering the NA on top of the same substrate (SiO$_2$) and surrounded by water. We design the resonance of the NA by controlling the geometrical parameters to match the resonance of the grating BIC 3'. The NA dimensions are width 30nm, thickness 30nm, total length 185m, and gap 15nm. Fig. 4(a) presents the intensity enhancement spectrum of the NA recorded at its gap center in the output plane (see the inset in Fi4. (a)). The resonant wavelength of about 910nm. We can observe that the NA has a low-quality factor Q =13. This large bandwidth ensures a weak coupling between the NA and the narrow linewidth BIC that guarantees a small shift of the BIC resonance after the coupling without substantial modifications of its properties [32].

**Hybrid system**

Last, we combined the slanted grating structure and the NA. Figs. 4 (a), (b), and (c) show the intensity enhancement spectra of respectively: the NA alone, the grating alone, and the entire structure and Fig. 4 (d), (e), and (f) show the intensity enhancement distributions respectively. Intensity enhancement is defined as the maximum intensity divided by the intensity of the incident field. We can observe that it has strong intensity enhancement at the center between the grating slabs and at the gap of the NA. For the case of NA alone, it is about 6000 around 910 nm. In the case of the grating only, the maximum intensity enhancement is 1400 around 910 nm. The entire structure (SBIC cavities + NA) has maximum intensity up to 1.6 x 10$^6$ around 920 nm. It is worth to mention that the intensity enhancement at the bottom is about half of the one at the center. Indeed, it seems that the higher intensity enhancement could be achieved by moving the NA to the center of the grating. However, in this case, it would affect the mode of the grating significantly and reduce the intensity enhancement eventually. In fact, even if the NA is placed out from the region where the slab resonance gives the maximum of intensity (at its center), the obtained enhancement of the hybrid structure is 3 orders of magnitude larger than the one of NA alone by keeping a very good localization in the gap region. The coupling efficiency is then estimated to 36% (slab enhancement of 700 at the NA position + NA enhancement of 6300 → 4.41x10$^6$ leading to 1.6x10$^6$/4.41x10$^6$=36,28%). This coupling efficiency is consistent with the value of 38% obtained in reference [4] but for a smaller enhancement factor of only 8000.

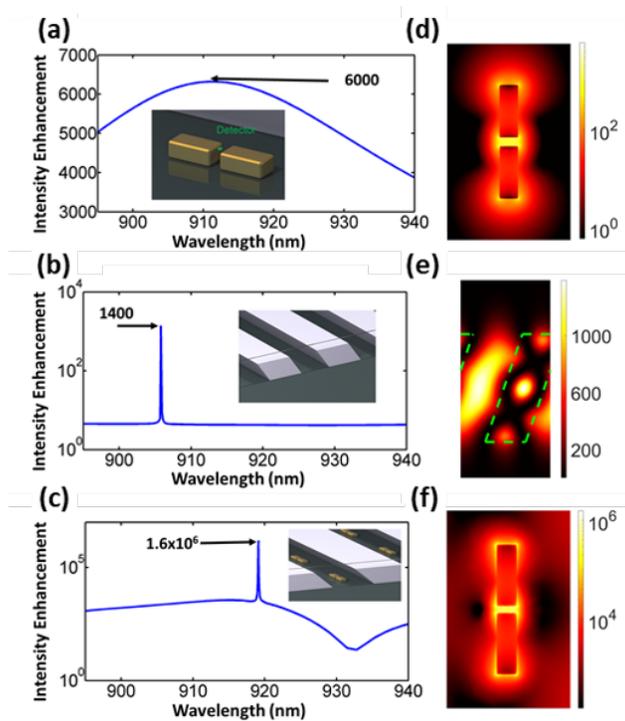

Fig. 4 Intensity enhancement spectra and intensity enhancement distributions. Intensity enhancement spectrum of (a) the NA. (b) the slanted grating. (c) the hybrid system. Intensity enhancement distribution of (d) the NA. (e) the slanted grating. (f) the hybrid system.

## Conclusion

We proposed and numerically demonstrated a new hybrid platform based on plasmonic optical nanoantenna and all dielectric grating. Our proposed platform exhibits an intensity enhancement of three orders of magnitude through a hybrid system made by plasmonic nano antenna and slanted Bound States in the continuum cavities.

This versatile platform can be used for applications requiring highly confined fields such as optical sensing, nonlinear optics, surface Raman spectroscopy, quantum optics, or for multiplex detection of antigens/antibodies, or biomarkers.

## Acknowledgments

The authors gratefully acknowledge the financial support of Boston University start-up funding.